\documentclass[twocolumn,aps,showpacs,preprintnumbers,amsmath,amssymb, superscriptaddress]{revtex4-1}
\pdfoutput=1

\usepackage{bm}
\usepackage{graphicx}
\usepackage{color}

\begin{document}

\title{Exotic Kondo crossover in a wide temperature region in the topological Kondo insulator SmB$_6$ revealed by high-resolution ARPES}

\author{N. Xu}

\email{nan.xu@psi.ch}

\affiliation{Swiss Light Source, Paul Scherrer Insitut, CH-5232 Villigen PSI,
Switzerland}

\author{C. E. Matt}

\affiliation{Swiss Light Source, Paul Scherrer Insitut, CH-5232 Villigen PSI,
Switzerland}

\affiliation{Laboratory for Solid State Physics, ETH Z\"urich, CH-8093 Z\"urich,
Switzerland}

\author{E. Pomjakushina}

\affiliation{Laboratory for Developments and Methods, Paul Scherrer Institut,
CH-5232 Villigen PSI, Switzerland}

\author{X. Shi}

\affiliation{Swiss Light Source, Paul Scherrer Insitut, CH-5232 Villigen PSI,
Switzerland}

\affiliation{Beijing National Laboratory for Condensed Matter Physics, and Institute
of Physics, Chinese Academy of Sciences, Beijing 100190, China}

\author{R. S. Dhaka}

\affiliation{Swiss Light Source, Paul Scherrer Insitut, CH-5232 Villigen PSI,
Switzerland}

\affiliation{Institute of Condensed Matter Physics, \'Ecole Polytechnique F\'ed\'crale de Lausanne, CH-1015 Lausanne, Switzerland}

\affiliation{Department of Physics, Indian Institute of Technology Delhi, Hauz Khas, New Delhi-110016, India}

\author{N. C. Plumb}

\affiliation{Swiss Light Source, Paul Scherrer Insitut, CH-5232 Villigen PSI,
Switzerland}

\author{M. Radovi\'c}

\affiliation{Swiss Light Source, Paul Scherrer Insitut, CH-5232 Villigen PSI,
Switzerland}

\affiliation{SwissFEL, Paul Scherrer Institut, CH-5232 Villigen PSI, Switzerland}

\author{P. K. Biswas}

\affiliation{Laboratory for Muon Spin Spectroscopy, Paul Scherrer Institut, CH-5232
Villigen PSI, Switzerland}

\author{D. Evtushinsky}

\author{V. Zabolotnyy}

\affiliation{Institute for Solid State Research, IFW Dresden, P. O. Box 270116, D-01171 Dresden,
Germany}

\author{J. H. Dil}

\affiliation{Institute of Condensed Matter Physics, \'Ecole Polytechnique F\'ed\'crale de Lausanne, CH-1015 Lausanne, Switzerland}

\affiliation{Swiss Light Source, Paul Scherrer Insitut, CH-5232 Villigen PSI,
Switzerland}

\author{K. Conder}

\affiliation{Laboratory for Developments and Methods, Paul Scherrer Institut,
CH-5232 Villigen PSI, Switzerland}

\author{J. Mesot}

\affiliation{Swiss Light Source, Paul Scherrer Insitut, CH-5232 Villigen PSI,
Switzerland}

\affiliation{Laboratory for Solid State Physics, ETH Z\"urich, CH-8093 Z\"urich,
Switzerland}

\affiliation{Institute of Condensed Matter Physics, \'Ecole Polytechnique F\'ed\'crale de Lausanne, CH-1015 Lausanne, Switzerland}

\author{H. Ding}

\affiliation{Beijing National Laboratory for Condensed Matter Physics, and Institute
of Physics, Chinese Academy of Sciences, Beijing 100190, China}

\affiliation{Collaborative Innovation Center of Quantum Matter, Beijing, China }

\author{M. Shi}

\email{ming.shi@psi.ch}

\affiliation{Swiss Light Source, Paul Scherrer Insitut, CH-5232 Villigen PSI,
Switzerland}

\date{\today}
\begin{abstract}

Temperature dependence of the electronic structure of SmB$_6$ is studied by high-resolution ARPES down to 1 K. We demonstrate that there is no essential difference for the dispersions of the surface states below and above the resistivity saturating anomaly ($\sim$ 3.5 K). Quantitative analyses of the surface states indicate that the quasi-particle scattering rate increases linearly as a function of temperature and binding energy, which differs from Fermi-Liquid behavior. Most intriguingly, we observe that the hybridization between the $d$ and $f$ states builds gradually over a wide temperature region (30 K $<$ T $<$ 110 K). The surface states appear when the hybridization starts to develop. Our detailed temperature-dependence results give a complete interpretation of the exotic resistivity result of SmB$_6$, as well as the discrepancies among experimental results concerning the temperature regions in which the topological surface states emerge and the Kondo gap opens, and give new insights into the exotic Kondo crossover and its relationship with the topological surface states in the topological Kondo insulator SmB$_6$.
\end{abstract}


\maketitle

\begin{figure*}[!t]
\begin{centering}
\includegraphics[width=4.5in]{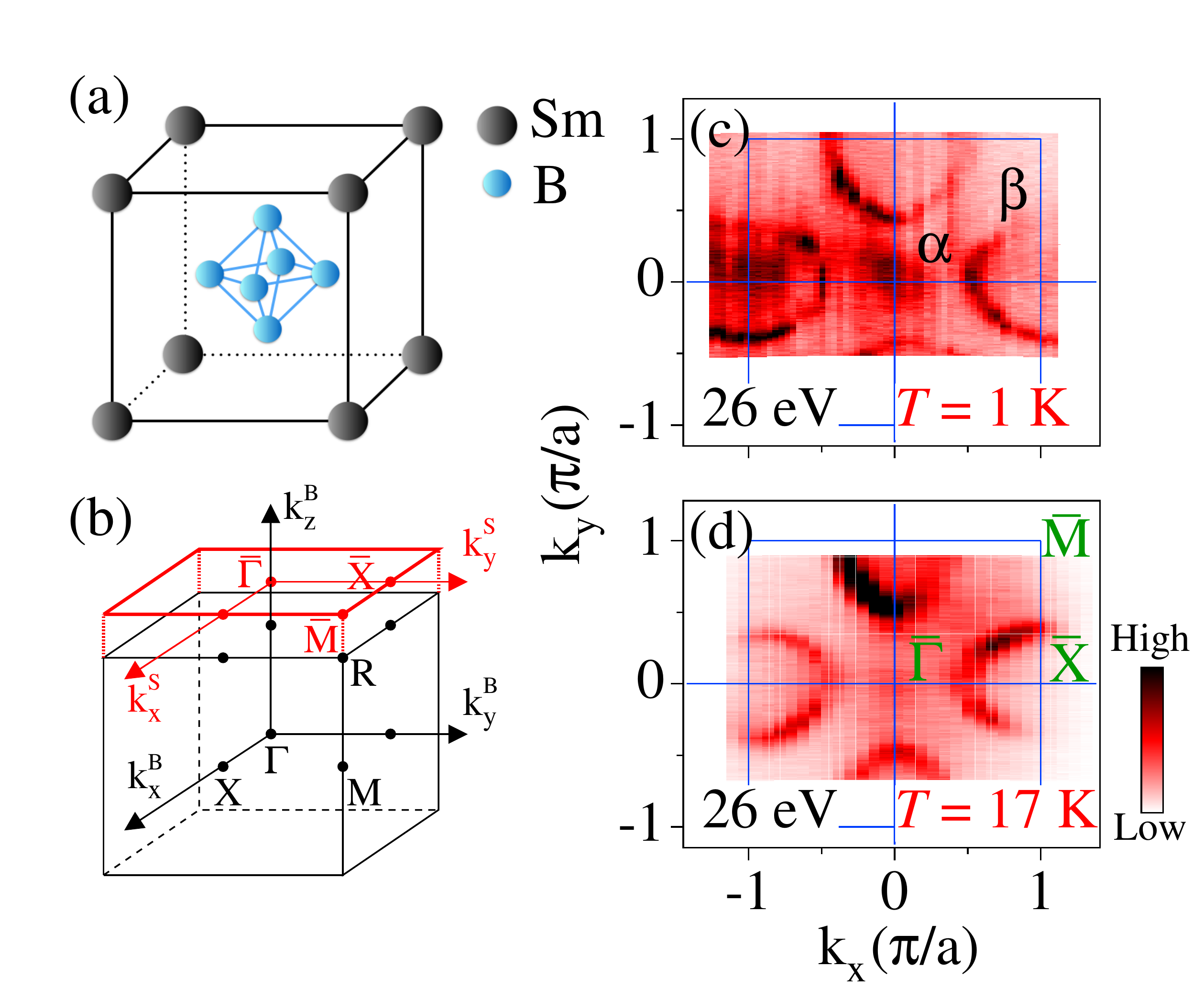} 
\par\end{centering}

\caption{\label{Fig1_FS} \textbf{Real- and momentum-space structure of SmB$_6$.} (a) CsCl-type structure of SmB$_6$ with Pm$\bar{3}$m space group. (b) First Brillouin zone of SmB$_6$ and the projection on the cleaving surface. High-symmetry points are also indicated. (c) ARPES intensity mapping of the Fermi surface at $T$ = 1 K for SmB$_6$ plotted as a function of 2D wave vector. The Fermi surface was measured at $h\nu$= 26 eV ($k_z$ = 0 for the bulk electron structure). The intensity is obtained by integrating the spectrum within $\pm$2 meV of $E_F$. (c) Same as (a), but for $T$ = 17 K. The intensity is obtained by integrating the spectrum within $\pm$5 meV of $E_F$.}

\label{Fig1_FS} 
\end{figure*}

\section{introduction}

Recently topological insulators (TIs) \cite{RMP_Hasan,RMP_SCZhang} with strong correlation effects have been extensively studied \cite{2010_DzeroPRL,2012_DzeroPRB,2013_FLuPRL,2010_Pesin_NP,2012_SCZhang_Sci,2013_XYDeng_PuB6}, and particular interest has focused on realizing new types of TIs in exotic materials that contain rare-earth elements. SmB$_6$, a well-known Kondo insulator (KI), has attracted much attention because it has been proposed to be a promising topological Kondo insulator (TKI) candidate where both electron correlations and nontrivial band topology play important roles  \cite{2010_DzeroPRL,2012_DzeroPRB,2013_FLuPRL}. At high temperature, SmB$_6$ behaves as a correlated bad metal. Upon decreasing temperature, a metal-to-insulator transition (MIT) occurs due to the opening of a hybridization band gap. However, below $\sim$ 3.5 K the resistivity saturates instead of diverging toward absolute zero, indicating the existence of in-gap states at low temperature \cite{1969_Menth_SmB6,1979_JAllen_SmB6,1995_Cooley_SmB6}. Strong evidence for surface-dominated transport at low temperature has been reported \cite{2013_wolgast_tranPRB,2013_Kim_tran_SP,2013_GLi_dHfA,2013_Kim_tran,2013_FChen_Tran,2013_ZYue_Tran,2013_Thomas_Than}, suggesting that the in-gap states have a surface origin. However, in transport measurements it is very challenging to distinguish the topology of these in-gap states. On the other hand, ARPES experiments have resolved that there is a surface Fermi surface (FS) formed by an odd number of electron-pockets around Kramers$'$ points in the surface Brillouin zone (SBZ) \cite{2013_Nan_TKI,2013_Neupane_SmB6,2013_DLFeng_SmB6}. Furthermore, a spin-resolved ARPES experiment has revealed that the metallic surface states are spin polarized, and the spin texture fulfils the condition that they are topologically nontrivial states protected by time reversal symmetry, thus indicating that SmB$_6$ is the first realization of a TKI \cite{2014_Nan_SpinSmB6}. However, so far, almost all the ARPES measurements have been carried out at temperatures near or above the resistivity anomaly ($\sim$ 3.5 K), below which the resistivity saturates. It is highly desirable to explore the detailed electronic structure and low-energy excitations of SmB$_6$ well below the resistivity saturating anomaly in order to understand its low-temperature electronic properties and to observe how such a topological state behaves in the presence of strongly correlation effects in the bulk. More fundamentally, there is controversy about the temperature behavior of the Kondo crossover and its relationship with the topological surface states \cite{2013_MYee_STM,2013_Nan_TKI,2013_Neupane_SmB6,2013_DLFeng_SmB6,2014_CMin_ARPES,2013_Johnson_2D3D}. In this letter we present high-resolution ARPES results obtained over a large temperature range from the high-temperature metallic phase down to very low-temperature (1 K), deep in the Kondo regime of the bulk states. The combination of very low sample temperatures, high energy-resolution ARPES, and high-quality SmB$_6$ single crystals make it possible to trace the detailed dispersions of the surface states in the narrow bulk band gap ($\sim$ 20 meV), as well as the spectral function of single-particle excitations as a function of temperature across the resistivity saturating anomaly. Our quantitative temperature-dependent ARPES results show that a crossover occurs from the high-temperature metallic phase to the low-temperature Kondo insulating phase over a wide temperature region. The data furthermore demonstrate how the topologically non-trivial surface states emerge. Our results give a comprehensive interpretation of the exotic resistivity behavior of SmB$_6$ in terms of the electronic structure and explain the discrepancies between various experimental results in the temperature region in which $d$-$f$ hybridization and topological surface states emerge.

High-quality single crystals of SmB$_6$ were grown by flux method. ARPES measurements were performed with VG Scienta R4000 electron analyzers using synchrotron radiation at the 1$^3$ endstation at BESSY and the SIS beamline at Swiss light source, Paul Scherrer Institut, with circular light polarization. The energy and angular resolutions were $\sim$  5 - 10 meV and 0.2$^{\circ}$, respectively. Samples were cleaved $in$-$situ$ along the (001) crystal plane in an ultrahigh vacuum better than 3$\times$10$^{-11}$ Torr. Shiny mirror-like surfaces were obtained after cleaving, confirming their high quality. The Fermi level of the samples was referenced to that of a gold film evaporated onto the sample holder.

\begin{figure*}[!t]

\begin{centering}
\includegraphics[width=5in]{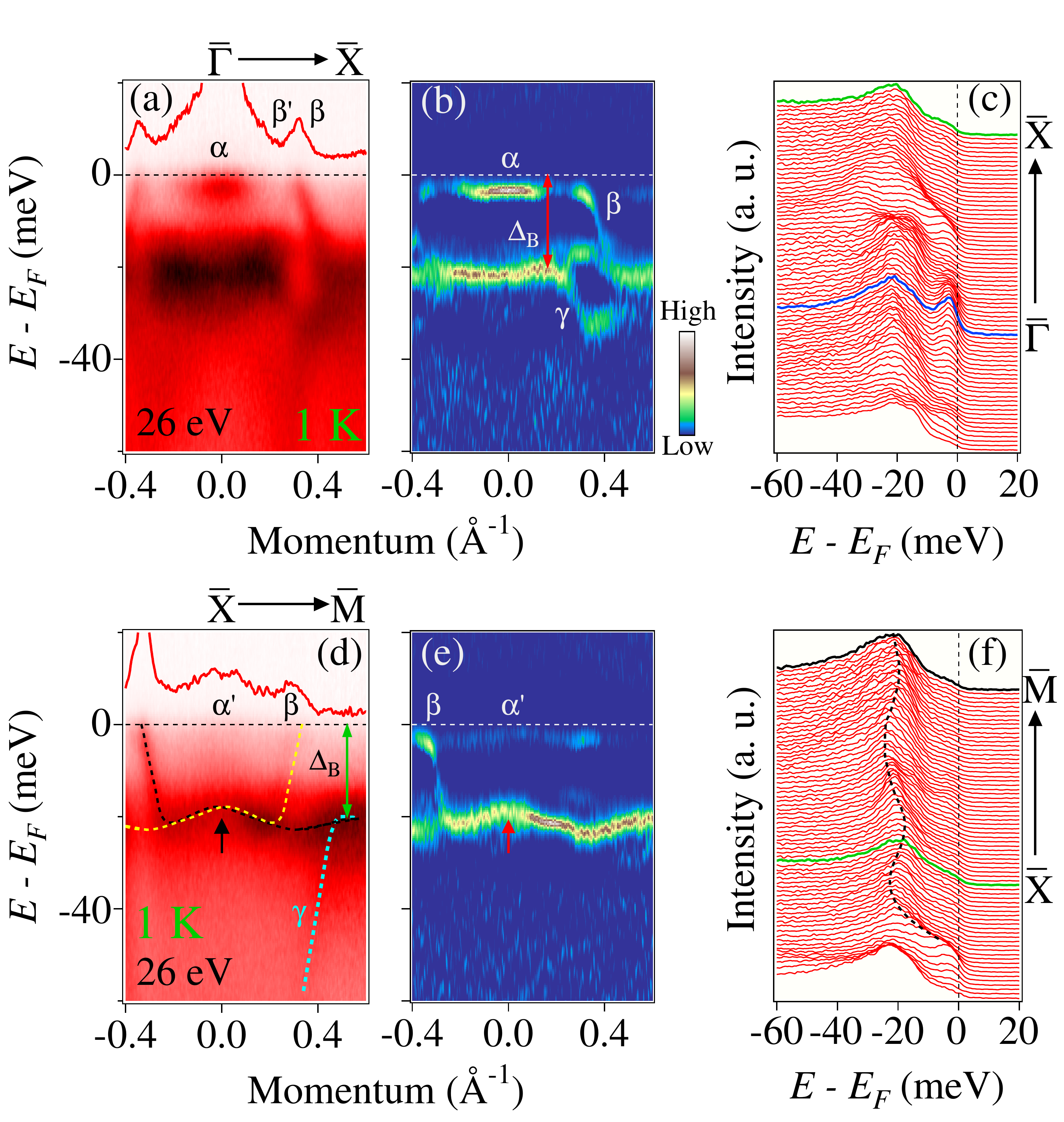} 
\par\end{centering}

\caption{\label{Fig2_FSandBS} \textbf{Surface band dispersions in SmB$_6$ at very low temperature (1 K).} (a) Near-$E_F$ ARPES intensity. (b) corresponding plot of the curvature of the EDC intensities. (c) Plot of EDCs measured at $h\nu$ = 26 eV as a function of wave vector and $E_B$ along the cut along the $\bar{\Gamma}-\bar{X}$ direction. The curve above is the MDC taken at $E_F$. (d)-(f) Analogous to (a)-(c), but for the cut along the $\bar{X}-\bar{M}$ direction. The dashed lines in (c) are the band dispersion obtained from the EDCs. The black (red) arrow in (c)/(d) indicates the crossing point of the two surface state branches with different spin-polarizations.}

\label{fig2_FSandBS} 
\end{figure*}

\section{result}
\subsection{Surface band structure below the resistivity saturating anomaly}

To compare the surface electronic structure at temperatures below and above the resistivity saturating anomaly ($\sim$ 3.5 K), we mapped the FS of the surface states at 1 K and 17 K. Figure \ref{Fig1_FS}(c) shows the FS mapped with h$\nu$ = 26 eV ($k_z$ = 4$\pi$ for the bulk states) at $T$ = 1 K, where the resistivity is fully saturated in transport measurements \cite{1969_Menth_SmB6,1979_JAllen_SmB6,1995_Cooley_SmB6}. The FS was obtained by integrating the ARPES intensity within a narrow energy window centered at $E_F$ ($\pm$ 2 meV). The definitions of the high symmetry points and their projections on SBZ of (100) surface are given in Fig. \ref{Fig1_FS}(b), which depicts the CsCl-type structure of SmB$_6$ in real space (Fig. \ref{Fig1_FS}(a)). For comparison, in Fig. \ref{Fig1_FS}(d) we plot the FS map at $T$ = 17 K \cite{2013_Nan_TKI}, which is well above the resistivity saturating anomaly. One can recognize that the topology of the FS is essentially the same on both sides of the resistivity saturating anomaly - namely, the FS is formed by the $\alpha$ pocket centered within the SBZ ($\bar{\Gamma}$ point) and the $\beta$ pockets sitting at the midpoints of the SBZ edges ($\bar{X}$ points). However the ARPES spectral weight of the FS is significantly enhanced at 1 K. This allows us to visualize the FS of the $\alpha'$ band at the $\bar{X}$ points (folding of the $\alpha$ band resulting from 1$\times$2 surface reconstruction) in addition to the $\beta'$ band observed in Ref. \cite{2013_Nan_TKI,2013_DLFeng_SmB6}. The identical dispersions of the surface states below and above the resistivity saturating anomaly revealed in our ARPES experiments confirm that the in-gap states inferred from transport measurements at very low temperature \cite{2013_wolgast_tranPRB,2013_Kim_tran_SP,2013_GLi_dHfA,2013_Kim_tran,2013_FChen_Tran} and surface states observed in the previous ARPES studies \cite{2013_Nan_TKI,2013_Neupane_SmB6,2013_DLFeng_SmB6} have essentially the same origin.

\begin{figure*}[!t]

\begin{centering}
\includegraphics[width=5in]{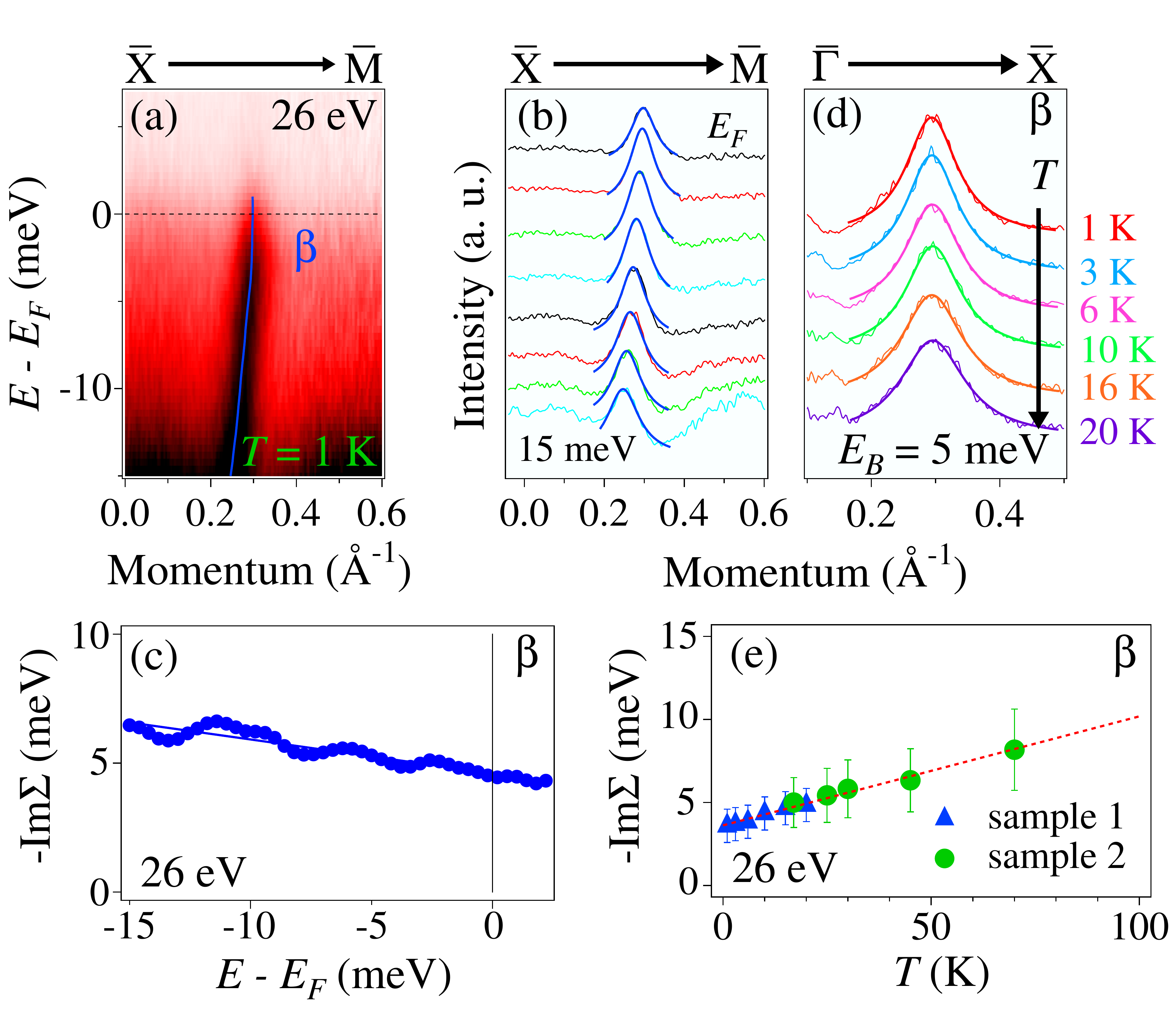} 
\par\end{centering}

\caption{\label{Fig3_selfenergy} \textbf{Quantitative analysis of the surface state dispersion in SmB$_6$.} (a) Near-$E_F$ ARPES intensity for SmB$_6$ measured at $h\nu$ = 26 eV as a function of $E_B$ and wave vector along the $\bar{X}-\bar{M}$ direction. The blue curve is the dispersion of the $\beta$ band traced by MDC fitting. (b) Corresponding MDC plots fitted by Lorentzian peaks (blue curves). (c) The imaginary part of the self-energy ($Im\Sigma$) of the Lorentzian-shaped MDC peaks at very low temperature (1 K). Standard deviations from the fitting are within 5$\%$ of the obtained value. (d) MDCs taken at $E_B$ = 5 meV with different temperatures, fitted by Lorentzian peaks (black curves). (e) Temperature dependence of $Im\Sigma$ of the Lorentzian-shaped MDC peaks at $E_B$ = 5 meV for two different samples.}

\label{Fig3_selfenergy} 
\end{figure*}

To trace the fine structure of the surface state inside the narrow Kondo gap, we carried out high-resolution ARPES measurement ($<$ 5 meV) at 1 K in order to suppress thermal broadening effects. Figures \ref{Fig2_FSandBS}(a) and \ref{Fig2_FSandBS}(d) display the intensity plots along the $\bar{\Gamma}$-$\bar{X}$ and $\bar{X}$-$\bar{M}$ directions, respectively. Similar to the observations at T $\ge$ 17 K \cite{2013_Nan_TKI}, the bulk $\gamma$ band hybridizes with the localized $f$ states, open a hybridization band gap $\Delta_B$ $\sim$ 20 meV, and this feature is enhanced in curvature plots \cite{P_Zhang_RSI2011} (Figs. \ref{Fig2_FSandBS}(b) and \ref{Fig2_FSandBS}(e)), as well as in energy distribution curves (EDCs) (Figs. \ref{Fig2_FSandBS}(c) and \ref{Fig2_FSandBS}(f)). Inside the hybridization band gap, we clearly see the $\alpha$ and $\beta$ bands centered at the $\bar{\Gamma}$ and $\bar{X}$ points, which form the FSs shown in Fig. \ref{Fig1_FS}(c). With the ultra-low temperature and high energy-resolution in the measurements, the spectral weight of the in-gap states is strongly enhanced, especially for the $\alpha$ band, which is clearly observed with a well-defined quasi-particle peak in the EDCs plot at 1 K (Fig. \ref{Fig2_FSandBS}(c)). The enhanced spectrum weight of the $\alpha$ band makes it possible to observe the folded $\alpha'$ band centered at  $\bar{X}$, as shown in the ARPES intensity plot and the momentum distribution curve (MDC) taken at $E_F$ (Fig. \ref{Fig2_FSandBS} (d)). We note that this $'$missing$'$ folding band has not been observed in previous ARPES experiments.

The high-quality data also enable us to trace the dispersion of the surface bands in detail. As shown in Fig. \ref{Fig2_FSandBS}(a), our results suggests a cone-like dispersion of the $\alpha$ band with a Dirac point (DP) very close to the bulk valence band. At deeper energies, the $\alpha$ band eventually merges with the localized $f$ states. For the $\beta$ band, due to photoemission matrix element effects, the left branch in Fig. \ref{Fig2_FSandBS}(d) is more enhanced than the right one. Following the dispersion of the $\beta$ band by fitting MDCs, we find that it linearly disperses from $E_F$ down to $E_B$ $\sim$ 20 meV and then shows a back-bending back behavior. The two branches cross each other at the $\bar{X}$ point as indicated by the black arrow, and finally merge to the bulk $f$ states as shown in Fig. \ref{Fig2_FSandBS}(d). This feature is better visualized in the curvature plot (Fig. \ref{Fig2_FSandBS}(e)) and in the EDC plots (Fig. \ref{Fig2_FSandBS}(f)). Thus, the ultra-low temperature and high-resolution ARPES results give a clear picture of the dispersions of the surface states, and show how these in-gap surface states, the $\alpha$ and $\beta$ bands, connect to the bulk valence bands. The new insights about the dispersion of the $\beta$ band at the $\bar{X}$ points naturally explains why its intensity suddenly decrease at a binding energy of $\sim$ 20 meV. It should be mentioned that a similar situation, but without a clear DP, is also observed on the first three-dimensional TI Bi$_{1-x}$Sb$_x$ \cite{2008_Hsieh_Nature}.

\begin{figure*}[!t]
\begin{centering}
\includegraphics[width=6in]{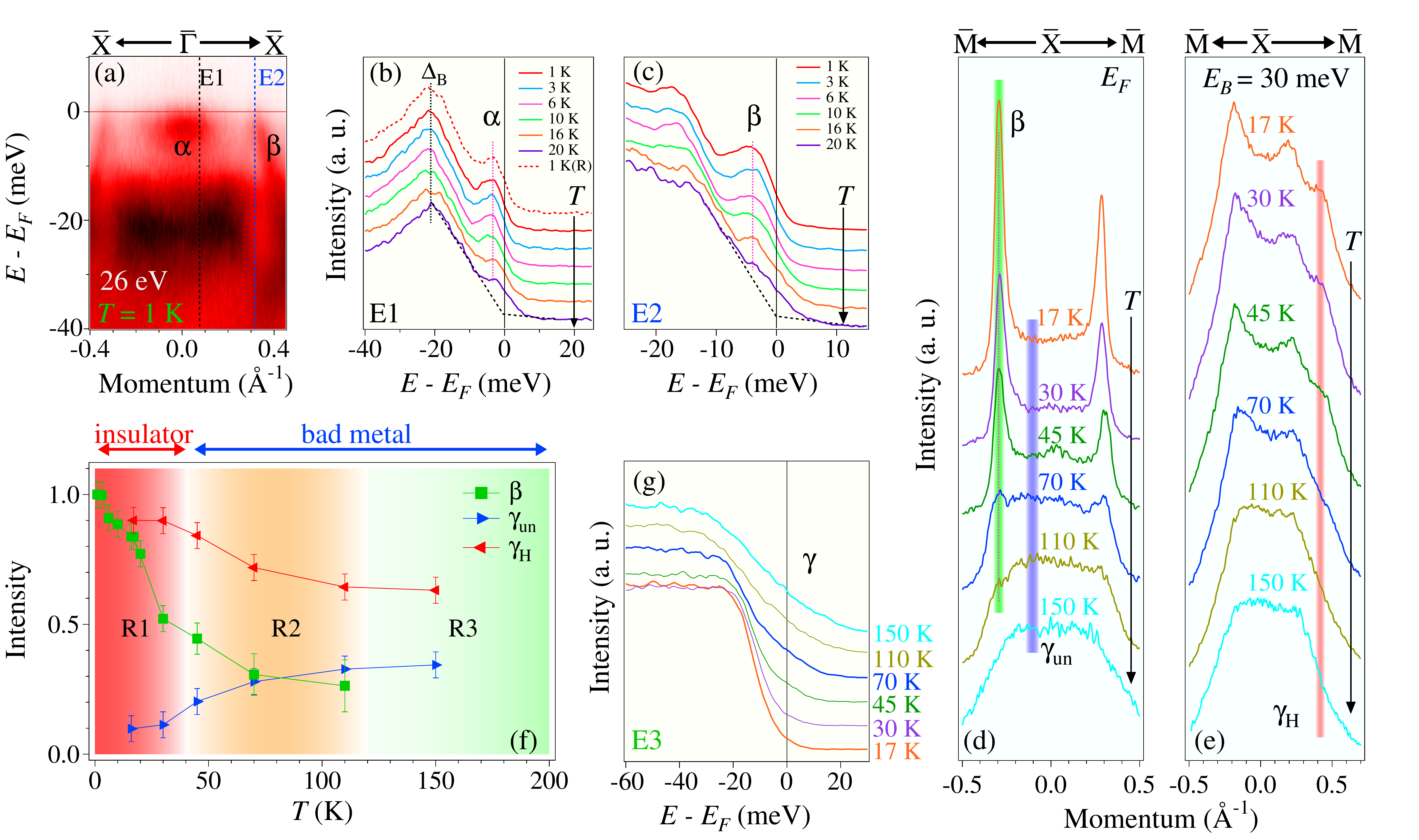} 
\par\end{centering}

\caption{\label{Fig4_temp} \textbf{Temperature dependence of the surface and bulk states in SmB$_6$.} (a) ARPES spectrum as a function of binding energy and wave vector along $\bar{X}$-$\bar{\Gamma}$-$\bar{X}$. (b)-(c), EDCs taken at the $k_F$ points of the $\alpha$ and $\beta$ bands (the EDCs indicated by E1 and E2 in (a)), respectively, at various temperatures. ARPES data taken after thermal cycling is shown by 1K(R) in (b), which demonstrates that the in-gap states are robust and protected against repeated thermal cycling. (d) MDCs taken at $E_F$ along the $\bar{X}$-$\bar{M}$ direction over the temperature range from 17 K to 150 K. (e) MDCs taken at $E_B$ = 30 meV along the $\bar{X}$-$\bar{M}$ direction over the same temperature range. (f) Temperature dependence of the intensity of the surface state $\beta$ band and the bulk conduction band with(without) hybridization with the $f$ electrons $\gamma_{H}$($\gamma_{norm}$). (g) EDCs taken at the $k_F$ points for the $\gamma$ band (the positions of the EDCs are indicated by E3 in Fig. \ref{Fig5_illustration}(a)), at various temperatures. }

\label{Fig4_temp} 
\end{figure*}

\subsection{Single particle scattering rate of the surface states}

To extract the single particle scattering rate of the surface states, we fit the MDCs with a single Lorentzian \cite{1999_TValla_Science}. The width of the Lorentzian peak, ${\Delta}k$($\omega$), is related to the quasi-particle scattering rate ${\Gamma}$($\omega$) = 2$|Im\Sigma(\omega)|$ = ${\Delta}k$($\omega$)$v_0$($\omega$), where $v_0$($\omega$) is the Fermi velocity and $|Im\Sigma(\omega)|$ is the imaginary part of the complex self-energy. Figure \ref{Fig3_selfenergy}(b) shows MDCs at various binding energies from the ARPES spectrum shown in Fig. \ref{Fig3_selfenergy}(a). The clean spectra at low binding energy near $E_F$ enable us to fit the MDCs accurately. However, at $E_B$ $>$ 15 meV, the spectra are mixed with the bulk states, as well as the bending back part of the $\beta$ band. As shown in Fig. \ref{Fig3_selfenergy}(c), the obtained $|Im\Sigma(\omega)|$ has a linear energy dependence that is not expected from the three-dimensional ($|Im\Sigma(\omega)| \propto \omega^2$) and two-dimensional ($|Im\Sigma(\omega)| \propto (\omega^2$/$\varepsilon_F$)ln(4$\varepsilon_F$/$\omega$)) Fermi liquid theory. On the other hand, this linear dependence of the scattering rate $\Gamma$($\omega$) in SmB$_6$ is similar to that of the Dirac fermions observed in graphene \cite{2007_Bostwick_Graphene}, and other 3D topological insulator, such as Bi$_2$Se$_3$ \cite{2012_ZHPan_PRL}. The unusual behavior of the energy dependence of the suppressed scattering rate suggests the topologically non-trivial nature of the surface state on SmB$_6$. We have also fitted the MDCs of the $\beta$ band at $E_B$ = 5 meV for different temperatures (Fig. \ref{Fig3_selfenergy}(d)) and obtained the temperature dependence of the self-energy as summarized in Fig. \ref{Fig3_selfenergy}(d). $Im\Sigma(\omega)$ increases with temperature at a constant rate, which also deviates from the Fermi liquid theory. It is worthy to mention the scattering rate increases faster than the previous non-interacting TIs \cite{2012_ZHPan_PRL} with rising temperature. This behavior may be due to the strong correlation effect in the TKI SmB$_6$ because electron-electron correlations can open additional channels to reduce the lifetime of single-particle excitations.

\begin{figure}[!t]
\begin{centering}
\includegraphics[width=3in]{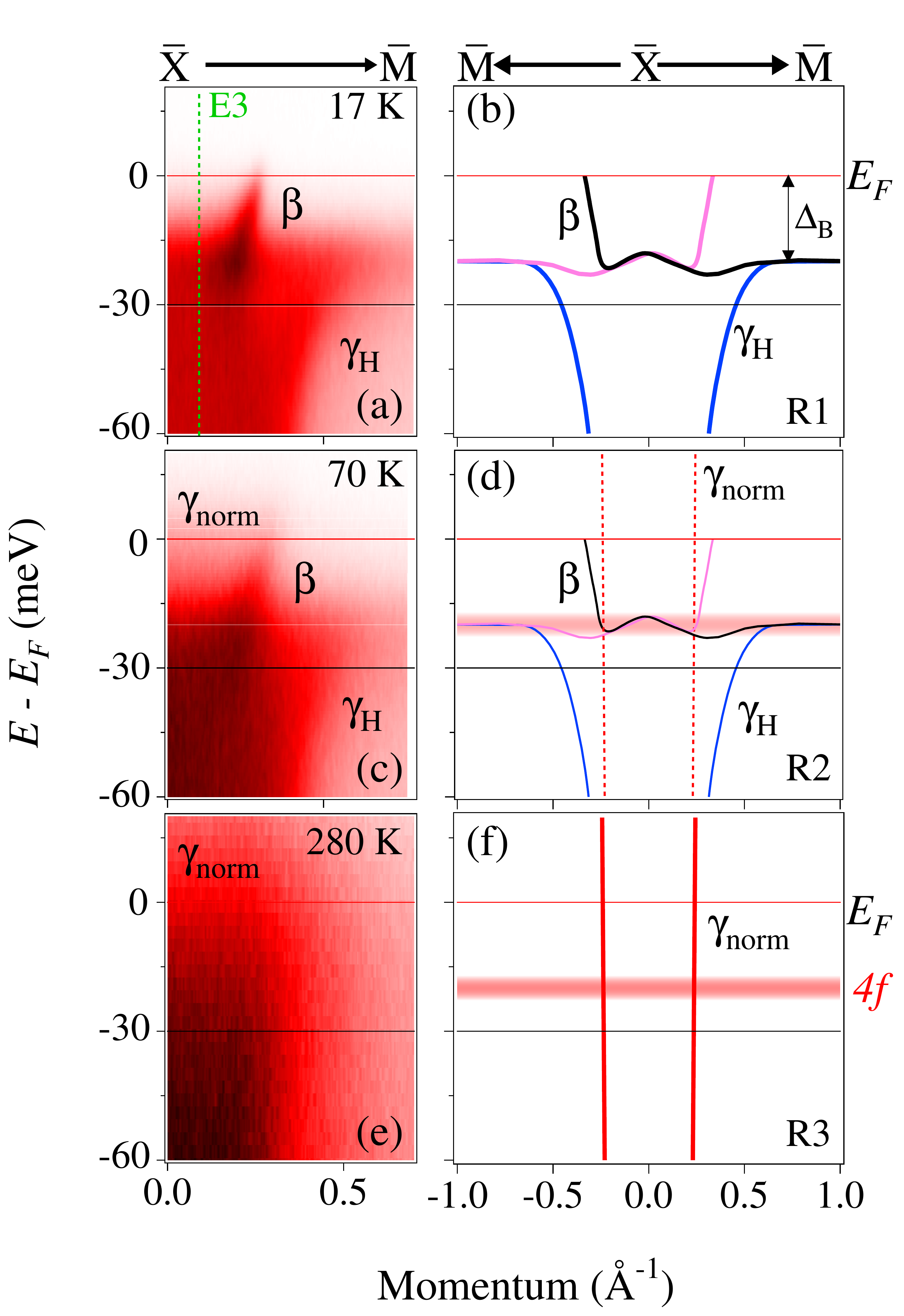} 
\par\end{centering}

\caption{\label{Fig5_illustration} \textbf{The illustrations of surface and bulk band structure at different temperature regions.} (a), (c) and (e), ARPES intensity plots for 17, 70, and 280 K, corresponding to temperature points in R1, R2, and R3 in Fig. \ref{Fig4_temp}(e), respectively. (b), (d) and (f) Illustrations of the band structure in R1, R2, and R3 in Fig. \ref{Fig4_temp}(f), respectively. }

\label{Fig5_illustration} 
\end{figure}

\subsection{Exotic Kondo crossover in a wide temperature region}

So far, the temperature behavior of the surface states and its relation to the Kondo gap arising from the $d$-$f$ hybridization in SmB$_6$ are still controversial issues. Some ARPES \cite{2013_Neupane_SmB6} and STM \cite{2013_MYee_STM} studies claim that the surface state exists inside the hybridization gap only below 30 K; above 30 K, the surface state disappears, accompanied by the complete destruction of the $d$-$f$ hybridization. On the other hand, other ARPES investigations on the same material \cite{2013_Nan_TKI,2013_DLFeng_SmB6,2014_CMin_ARPES} indicate that the surface states can exist as high as 110 K. The interplay between the surface states and the hybridization between the $\gamma$ band and $f$ states, as well as their relationship with the exotic resistivity-temperature behavior \cite{1969_Menth_SmB6,1979_JAllen_SmB6,1995_Cooley_SmB6} are currently under debate. To examine the temperature dependence and evolution of the surface states and the $d$-$f$ hybridization, we performed detailed high-resolution ARPES measurements in the temperature range from 1 K to 280 K. Figure 4(b) shows the evolution of the EDC located at the $k_F$ of the $\alpha$ band, as indicated by E1 in Fig. \ref{Fig4_temp}(a), from 1 K to 20 K. The $\alpha$ band has a well-defined quasi-particle peak in the temperature region below 3.5 K where the resistivity saturates. Upon increasing temperature, the coherent spectral weight decreases monotonically but does not show any anomaly up to 20 K. Cooling the sample back to 1 K, the spectral weight is recovered without any degradation caused by aging (1K(R) in Fig. \ref{Fig2_FSandBS}(b)). A similar temperature dependence is also observed for the $\beta$ band in Fig. \ref{Fig4_temp}(c) (E2 cut in Fig. \ref{Fig4_temp}(a)). Above 20 K, it is difficult to trace the quasi-particle peaks in the EDCs for both $\alpha$ and $\beta$ due to thermal broadening of the strong $f$ electron peak at $E_B \sim$ 20 meV. However, the spectral peak can still be clearly observed in the MDC at $E_F$, as shown in Fig. \ref{Fig4_temp}(d) for temperatures from 17 K to 150 K. The double peaks of the $\beta$ band monotonically become weaker with increasing temperature and disappear between 110 K and 150 K. Our results demonstrate that the surface states can exist up to temperatures as high as 110 K. However, we notice that at temperatures T $\ge$ 45 K, some intensity emerges between the two MDC peaks. The spectral weight in this region increases as the temperature is raised and becomes dominant at 150 K, corresponding to the bulk band ($\gamma_{norm}$), which crosses $E_F$ without hybridizing with the $f$ states. The temperature behavior of the $\gamma$ band is confirmed by the EDCs taken at the $k_F$ of the $\gamma_{norm}$ band for different temperatures in Fig. 4(g), with the position indicated by E3 in Fig. \ref{Fig5_illustration}(a). When T $\ge$ 45 K, some intensity emerges within the hybridization gap ($E_B <$ 20 meV) and becomes dominant at T $>$ 110 K. 
In order to quantitatively analyze the strength of the $d$-$f$ hybridization, we plot the MDCs for different temperatures at $E_B$ = 30 meV in Fig. \ref{Fig4_temp}(e), with the position indicated by the black line in Fig. \ref{Fig5_illustration}(a)-\ref{Fig5_illustration}(f). Besides the double peaks centered around the $\bar{X}$ point, which are the residual intensity of the $\beta$ band surface state, an additional peak on the right shoulder corresponds to the $\gamma_H$ band which hybridizes with the $f$ states (see Figs. \ref{Fig5_illustration}(a) and \ref{Fig5_illustration}(b)), as indicated by the red region in Fig. \ref{Fig4_temp}(e). The intensity of the $\gamma_H$ band shows opposite temperature behaviour to the $\gamma_{norm}$ band, decreasing as temperature is raised.
We extract the intensity for the surface band $\beta$ (peak intensity in the green region in Fig. \ref{Fig4_temp}(d)) and the bulk band $\gamma_{norm}$, which crosses $E_F$ without hybridizing with the $f$ electrons (intensity in the blue region in Fig. \ref{Fig4_temp}(d)), as well as the intensity of the bulk band $\gamma_H$, which hybridizes with the $f$ electrons (intensity in the red region Fig. \ref{Fig4_temp}(e)). These intensities are plotted for different temperatures in Fig. \ref{Fig4_temp}(f). Our temperature results suggest that the hybridization between the $\gamma$ band and localized $f$ states is a crossover process in a large temperature region. At the $d$-$f$ hybridization crossover region (R2 in Fig \ref{Fig4_temp}(f)), the $\gamma_{norm}$ band crossing $E_F$ without hybridization with the $f$ states coexists with the hybridized $\gamma_H$ band. The ratio of $\gamma_{norm}$/ $\gamma_{H}$ decreases with cooling down the temperature, and reach the minimum at $T <$ 30 K. This indicates the Kondo crossover in SmB$_6$ occurs over a wide temperature range, starting at $T =$ 110 K and completing at $T =$ 30 K.

\section{discussion}

The aforementioned results, especially the observation of the Kondo crossover in a wide temperature region, provide new insights into the evolution of the electronic structure with temperature in SmB$_6$ and its connection with the exotic resistivity as a function of temperature \cite{1969_Menth_SmB6,1979_JAllen_SmB6,1995_Cooley_SmB6}. In the high temperature region (R3 in Fig. \ref{Fig4_temp}(f)), the bulk $\gamma_{norm}$ band crosses $E_F$ without hybridizing with the $f$ states, as illustrated in Fig. \ref{Fig5_illustration}(f), as well as in the ARPES intensity plot at 280 K in Fig. \ref{Fig5_illustration}(e). Therefore, in this temperature regime, SmB$_6$ shows metallic behavior in transport due to the carriers contributed by the FSs of the $\gamma_{norm}$ band. When the material is cooled gradually from 110 K to 45 K (R2 in Fig. \ref{Fig4_temp}(f)), the bulk $\gamma_{norm}$ band starts to hybridize with the $f$ states, as indicated by the bent back dispersion at $E_B \sim$ 20 meV shown in Figs. \ref{Fig5_illustration}(c)-\ref{Fig5_illustration}(d). During this crossover region, the partial $\gamma_{norm}$ band still crosses $E_F$, coexisting with the hybridized $\gamma_{H}$ band, with the ratio of relative intensities $\gamma_{norm}$/$\gamma_{H}$ decreasing with falling temperature (Fig. \ref{Fig4_temp}f). In the meantime, the surface state emerges when the $d$-$f$ hybridization starts to develop and band inversion of the $f$ and $d$ electrons occurs. Due to the partial $\gamma_{norm}$ band that still crosses $E_F$ in the crossover region (R2), SmB$_6$ is still a bad metal in this region. When SmB$_6$ is cooled further down to below 30 K, the hybridization between the $\gamma$ band and the $f$ states becomes complete, as indicated by a clean gap opening at the $k_F$ position of the $\gamma$ band in Fig. \ref{Fig4_temp}(g). The complete bulk hybridization and gap opening (as seen in Figs. \ref{Fig5_illustration}(a)-\ref{Fig5_illustration}(b)) turns the system into a bulk insulator and corresponds to the MIT transition \cite{1969_Menth_SmB6,1979_JAllen_SmB6,1995_Cooley_SmB6}. At very low temperatures (below 3.5 K), the surface states dominate the transport properties, causing the resistivity to saturate instead of diverging as the temperature approaches zero. 

Our temperature-dependent data unify the seemingly conflicting observations on SmB$_6$ by different groups. In the crossover region (R2 in Fig. \ref{Fig4_temp}(f)), the weak surface state $\alpha$ and $\beta$ bands can hardly be distinguished from the tail of the broad and strong $f$ states peak sitting at a shallow binging energy of about 20 meV. As a result, the surface state bands are only observed in the very near-$E_F$ MDCs measured by ARPES \cite{2013_Nan_TKI,2013_DLFeng_SmB6,2014_CMin_ARPES}. On the other hand, in the temperature crossover region, due to the existence of the partial $\gamma_{norm}$ band crossing $E_F$, the gap seems closed as observed in density of states for both partially angle-integrated ARPES \cite{2013_Neupane_SmB6} and STM \cite{2013_MYee_STM}. 
Our temperature dependent ARPES results on SmB$_6$ give a comprehensive picture of the development of the topological surface states and the Kondo gap due to the $d$-$f$ hybridization, which could account for its exotic resistivity behavior as a function of temperature. To our knowledge, this constitutes the first observation that the Kondo crossover in SmB$_6$ takes place over such a wide crossover temperature regime, and the origin of such behavior deserves further studies. One mechanism candidate is that the Kondo temperature near the surface region is different from the one in the bulk, following from the fact that the conduction electrons density of states at the surface differ from that in the bulk.

\section*{ACKNOWLEDGMENTS}

We acknowledge H.M. Weng, X. Dai and Z. Fang for helpful discussions. This work was supported by the Sino-Swiss Science and Technology Cooperation (Project No. IZLCZ2138954), the Swiss National Science Foundation (No. 200021-137783), and MOST (2010CB923000), NSFC.

\bibliography{TKI_long}

\end{document}